\documentclass[12pt]{article}
\usepackage[utf8]{inputenc}
\usepackage{amssymb}
\usepackage{alphabeta}
\usepackage{amsmath}
\usepackage{graphicx}
\usepackage{cancel}
\graphicspath{ {./images/} }
\newcommand{\iprod}{\mathbin{\lrcorner}}
\usepackage{amsthm}
\usepackage{amsmath}
\usepackage{blindtext}
\usepackage{geometry}
\title{A geometric framework for conservation laws along null hypersurfaces}
\author{Eva Politou\\
\\
University of Toronto, Department of Mathematics}

\vspace{0.4cm}
\date{\today}

\begin{document}

\maketitle
\begin{center}
    \textbf{Abstract}\\
\end{center}
 We use the general theory of local conservation laws for arbitrary partial differential equations to provide a geometric framework for conservation laws on characteristic null hypersurfaces. The operator of interest is the wave operator on general four-dimensional Lorentzian manifolds restricted on a null hypersurface.
 
\tableofcontents

\pagebreak
 \section{Introduction}
 
 \subsection{Motivation}

In \cite{Aretakis17} Aretakis found the necessary and sufficient conditions for the existence of conservation laws on null hypersurfaces for the wave equation on four-dimensional Lorentzian spacetimes using standard geometrical analysis. The existence of conservation laws in spacetimes such as the Minkowski space, extremal black holes and the null infinity of asymptotically flat spacetimes has been extensively studied by Aretakis \cite{Aretakis12},\cite{Aretakis17}, Reall and Lucietti \cite{LucRea12}, and Murata \cite{Murata13}. \\
\\
In this paper, we use the general theory of local conservation laws for arbitrary PDEs in order to provide a geometric derivation of the conservation laws on characteristic hypersurfaces.\\
\\
The generalization of Noether's 1st theorem as can be found in  \cite{Khavkine15} and \cite{Olver86} states that conservation laws are in bidirectional correspondence with what might as well be called co-symmetries. Symmetries lie in the kernel of the PDE operator. On the other hand, co-symmetries lie in the kernel of the formal adjoint operator,  PDE*, of the original PDE operator. Proofs of this theorem can be found in \cite{glenn},\cite{Krasil98} and \cite{Verbo98}. The generalized Noether's 2nd theorem states that the formal adjoint is the same as the operator itself for equations in Euler-Lagrange form, so that symmetries and co-symmetries coincide, thus restoring the classical situation.\\
\\
In this paper we examine the wave operator restricted on a null hypersurface of a four-dimensional Lorentzian manifold. The dependent variables are two scalar functions, the field itself and its transverse null derivative. Since there is only one scalar constraint
equation, we define its formal adjoint acting on a single scalar function and returning a pair of functions. \\
\\

\subsection{Geometric Setting} 
In this paper, we use the geometric setting of double null foliation. Considering $\mathcal{H}$ to be a regular null hypersurface of a four-dimensional Lorentzian manifold $(M, g)$ and ${S} = {S}_v$ a foliation of $\mathcal{H}$ where the sections vary smoothly in $v \in \mathcal{R}$, then $\mathcal{H}= \cup_v {S}_v$ can be uniquely determined by the choice of one section ${S}_0$, the choice of a smooth function $\Omega$ on $\mathcal{H}$ and the choice of a null geodesic vector field $L_{geod}$ tangential to the null generators of $\mathcal{H}$ such that $\nabla_{ L_{geod}}L_{geod} = 0$.\\
\\
Then we consider the null vector field, $\underline{L}_{geod}$, normal to the sections $S_v$ and conjugate to $\mathcal{H}$ such that it satisfies the relation
$$g( L_{geod},\underline{L}_{geod})=-\Omega^2.$$
We take $u$ and $v$ to be the optical functions that satisfy
$$\nabla u=-L_{geod},\;\;\;\;\;\nabla v=\underline{L}_{geod}.$$
Then by constructing a diffeomorphism $Φ_{u,v}$ from any sphere ${S}_{u,v}$ to ${S}_0$ and a diffeomorphism $Φ$ from ${S}_0$ to the unit sphere  $\mathbf{S}_2$ we get a diffeomorphism from the spheres  ${S}_{u,v}$ to $\mathbf{S}_2$.\\
\\
In the double null coordinate system $(u,v,\theta^1,\theta^2)$ where $(\theta^1,\theta^2)$ is an arbitrary local coordinate system on the spheres $S_{u,v}$, the wave equation restricted on the null hypersurface $\mathcal{H}$ takes the form
$$-2 \partial _v \partial _u (\sqrt[4]{\cancel{g}}ψ)+\sqrt[4]{\cancel{g}}\cdot Q^Sψ=0.$$
Here  $\cancel{g}$ is the induced metric on the 2-spheres $S_{u,v}$ and $Q^S:C^{\infty}(Η)\rightarrow \mathcal{R}$ is an operator
acting on  a solution to the wave equation $ψ$,  given by:
    $$Q^Sψ=Ω^2 \cancel{\Delta} ψ+(\cancel{\nabla}Ω^2 -2Ω^2ζ^{\#})\cancel{\nabla}ψ+ω ψ$$
where $$ω= \partial_v(Ωtr\underline{χ})+\frac{1}{2}(Ωtr\underline{χ})(Ωtrχ)$$
and $ζ^\#$, $χ$ and $\underline{χ}$ are given below. Note that $\#$ refers to the sharp musical isomorphism.\\
\\
In a normalized frame $(e_1,e_2,e_3,e_4 )$ where $\{e_1,e_2\}=(e_A)_{A=1,2}$ is an arbitrary frame on the spheres $S_{u,v}$, $e_3 =Ω \underline{L}_{geod}$ and $e_4 =Ω L_{geod}$, the torsion $ζ^\#$ is defined as
$$ζ_{Α} :=(\nabla _A e_4, e_3)$$
and $χ$ and $\underline{χ}$ the null second fundamental forms are defined as follows
$$χ_{ΑΒ}:=g(\nabla_A e_4,e_B)$$
$$\underline{χ}_{ΑΒ}:=g(\nabla_A e_3,e_B)$$
with 
$$trχ:=\cancel g^{AB}χ_{ΑΒ},\;\;\;\;\;\;tr\underline{χ}:=\cancel g^{AB}\underline{χ}_{ΑΒ}$$
\\
For any section ${S}_v$, there exist a unique metric, $\hat{g}$, conformal to the induced metric on  ${S}_v$, $\cancel{g}$, such that the volume form of the conformal metric is equal to the volume form of the unit metric on the sphere $g_\mathbf{S^2}$.
We also introduce the conformal factor $\phi$ such that $\cancel{g}=\phi^2\cdot \hat{g}$. Then, the wave equation becomes
$$-2\partial_u \partial_v (\phi \cdot ψ)+\phi \cdot Q^sψ=0.$$

 \subsection{Literature}

 In \cite{Aretakis17} Aretakis showed that  a null hypersurfaces admits a first order conservation laws with respect to a foliation $S$ iff the linear space 
 \begin{equation*}
     \mathcal{W}^S=\{\Theta^S \in C^{\infty}(\mathcal{H}): \Omega^2 L_{geod}\Theta^S=0, \;\; \partial_v\Big(\int_{S_v}Y^S(\phi \cdot ψ)\cdot \Theta^Sdμ_{\mathbin{S^2}}\Big)=0\}
 \end{equation*}
 has 
 \begin{equation*}
     dimW^S\geq 1.
 \end{equation*}
Here $Y^S$ is the unique null vector which is normal to the sections $S_v$, conjugate to $\mathcal{H}$ and satisfies $g(L_{geod},Y^S)=-1$. Then he defines $W^S$ and $dimW^S$ as the kernel and dimension of the conservation law.
\\
\\
In the same paper, Aretakis proves that a null hypersurface $\mathcal{H}$ admits a conservation law with respect to a foliation $S$ in the sense of the definition above iff there is a non trivial linear space $U^S \subset V^{\mathcal{H}}=\{f \in C^{\infty}(\mathcal{H}): \Omega^2L_{geod}=0\}$ such that
\begin{equation*}
    O^S\big(\frac{1}{\phi}\Theta^S\big) =0 \;\;on\;\; \mathcal{H}, \; for\; all\; \Theta^S \in U^S
\end{equation*} 
where $O^S$ is defined to be the adjoint of $Q_v$ given by \\
    $$O^Sψ=Ω^2 \cancel{\Delta} ψ+(\cancel{\nabla}Ω^2 +2Ω^2 ζ^{\#}) \cancel{\nabla}ψ+ \left[ 2\cancel{{div}}(Ω^2\cdot ζ^{\#})+\partial_v(Ωtr(\underline{χ})+\frac{1}{2}(Ωtr(\underline{χ})(Ωtr(χ)\right]ψ.$$ 
\\

\section{The Main Theorem}

For the purposes of this paper we use the following definition for the formal adjoint of an operator.\\
\\
\textbf{Definition}: Let $M$ be a compact manifold with boundary. If we have two vector bundles $E,F \rightarrow M$ with inner products and a differential operator $D:C^{\infty}(E)\rightarrow C^{\infty}(F)$ then $D$ admits a formal adjoint $D^*:C^{\infty}(F) \rightarrow C^{\infty}(E)$. For a given operator $D$, we define its \textbf{adjoint} $D^*$ such that for smooth sections $ψ\in C^{\infty}(E)$ and $f \in C^{\infty}(F)$ it satisfies the following identity:
\begin{equation}
\label{eqn:def}
    \big< D[ψ],f\big> dvol -\big< ψ, D^*[f]\big> dvol = dW[ψ,f]
\end{equation}
where d is the exterior derivative,  $W[ϕ,ψ]$ is called the bilinear concomitant and is valued in $(dimM-1)$-forms. Variations of this definition can be found in \cite{Courant55} and \cite{Zwillinger57}.\\
\\
The main result of the present paper is the following:\\

\textbf{Theorem:} Let $\mathcal{H}$ be a regular null hypersurface of a four-dimensional Lorentzian manifold $(\mathcal{M},g)$. Let $D$ be the wave operator restricted on $\mathcal{H}$ and $D^*$ its associated adjoint, defined by 
\begin{equation*}
    \big< D[(ψ, \partial_u (ϕ\cdot ψ))],f\big>\epsilon -\big< (ψ, \partial_u (ϕ\cdot ψ)), D^*[f]\big>  \epsilon = dW[ψ,\partial_u (ϕ\cdot ψ),f],
\end{equation*}
 where $ψ$, $f$ smooth functions on $\mathcal{H}$, $\epsilon$ the volume form of $\mathcal{H}$, and $W$ the 2-form tri-linear concomitant. Then, $W[ψ,\partial_u (\phi\cdot ψ),f]$ is given by
       \begin{equation*}
           W[ψ,\partial_u (ϕ\cdot ψ),f]=\frac{-2f\partial_u (\phi \cdot ψ)}{\phi}\cancel{\epsilon}+\epsilon \iprod (fΩ^2\cancel{\nabla}ψ-Ω^2ψ\cancel{\nabla}f-fψ2Ω^2 ζ^{\#})
       \end{equation*}    
with $\cancel{\epsilon}$  the induced volume form on the sections $S_v$.
Moreover, if $ψ$ is a solution to the wave equation such that $D[ψ]=0$ and $f$ a smooth function such that $D^*[f]=0$ then, $dW=0$ and $W$ is a closed 2-from and hence the integrals 
\begin{equation*}
    \int _{S_v}W
\end{equation*}
are conserved, that is independent of $v$. 

\subsection*{Example}
We present here the example of Minkowski spacetime. In the geometric setting of the double null foliation, in the double null coordinates $(u,v,\theta^1,\theta^2)$, the wave operator restricted on $\mathcal{H}$ takes the form
\begin{equation*}
    D[ψ, \partial_u (r\cdot ψ)]=-\frac{2}{r}\partial_v (\partial_u (r \cdot ψ))+ \cancel{Δ}ψ
\end{equation*}
where 
$\phi=r$, $\Omega=1$,$trχ=\frac{2}{\sqrt{2}r}$, $tr\underline{χ}=-\frac{2}{\sqrt{2}r}$ and $ζ^{\#}=0$ so, $w=0$.
\\
Then,
\begin{equation*}
   D^* [f]=\bigg(\cancel{Δ}f,\frac{2}{r^2}\partial_v (r \cdot f)\bigg),
\end{equation*}

\begin{equation*}
    W[ψ,\partial_u (ϕ\cdot ψ),f]|_{\partial H}= \frac{h}{ϕ^2}\cancel{\epsilon}|_{\partial H}=\frac{h}{r^2}\cancel{\epsilon}=\frac{ -2f\partial_u (r \cdot ψ)}{r}\cancel{\epsilon}.
\end{equation*}
\\
In order to have a conservation law we need $D^*[f]=0$. Indeed for $f=\frac{1}{r}$ we get $D^*[\frac{1}{r}]=0$ which then gives 
\begin{equation*}
    W[ψ,\partial_u (ϕ\cdot ψ),f]=\frac{ -2\partial_u (r \cdot ψ)}{r^2}\cancel{\epsilon}=-2\partial_u (r \cdot ψ))dμ_{\mathbin{S}^2},
\end{equation*}
leading to the same conservation law as in \cite{Aretakis17} \begin{equation*}
  \int _{S_v}-2\partial_u (r \cdot ψ))dμ_{\mathbin{S}^2}.
\end{equation*}

\section{Proof}

\textbf{Proof of Theorem}: Let $D:C^{\infty}(\mathcal{H}) \times C^{\infty}(\mathcal{H}) \rightarrow C^{\infty}(\mathcal{H})$ be the restricted wave operator  on $\mathcal{H}$ taking a pair of functions $(ψ, \partial_u (\phi\cdot ψ))$ on the null hypersurface $\mathcal{H}$ into the scalar function
\begin{equation}
\label{eqn:doper}
D[ψ, \partial_u (\phi\cdot ψ)]=-\frac{2}{\phi}\partial_v (\partial_u (\phi \cdot ψ))+Ω^2 \cancel{Δ}ψ +(\cancel{\nabla}Ω^2-2Ω^2 ζ^{\#})\cancel{\nabla}ψ+wψ
\end{equation}
where on $\mathcal{H}$
$$w=\bigg[ \partial_v (Ω tr\underline{χ})+\frac{1}{2}(Ω trχ)(Ω tr\underline{χ}) \bigg].$$
\\
\\
We are looking for its formal adjoint, $D^*:C^{\infty}(\mathcal{H})  \rightarrow C^{\infty}(\mathcal{H})\times C^{\infty}(\mathcal{H})$ which will take a function $f$ into a pair of functions
$$D^* [f]=(a,b)$$
where $a,b$ scalar functions. Then, equation (\ref{eqn:def}) takes the form 
\begin{equation}
\label{eqn:basiceq}
    \big< D[(ψ, \partial_u (\phi\cdot ψ))],f\big>\epsilon -\big< (ψ, \partial_u (\phi\cdot ψ)), D^*[f]\big>  \epsilon = dW[ψ,\partial_u (\phi\cdot ψ),f].
\end{equation}
\\
Thus, in the case of the wave operator restricted on a null hypersurface, the first term on the left hand side of equation (\ref{eqn:basiceq}) looks like
\begin{equation}
\begin{split}
      \big< D[(ψ, \partial_u (\phi\cdot ψ))],f\big> \epsilon &= \big< -\frac{2}{\phi}\partial_v (\partial_u (\phi \cdot ψ)+Ω^2 \cancel{Δ}ψ +(\cancel{\nabla}Ω^2-2Ω^2 ζ^{\#})\cancel{\nabla}ψ+wψ,f\big> \epsilon\\
&= \Big(-\frac{2}{\phi}\partial_v (\partial_u (\phi \cdot ψ)+Ω^2 \cancel{Δ}ψ +(\cancel{\nabla}Ω^2-2Ω^2 ζ^{\#})\cancel{\nabla}ψ+wψ\Big)f \epsilon  
\end{split}
\end{equation}
where the volume form in this case is $ \epsilon  =dvdμ_{g\mkern-7.5mu/}$.\\
\\
Since $D^*[f]=(a,b)$, the second term of the right hand side of the equation (\ref{eqn:basiceq}) looks like
\begin{equation}
    \begin{split}
        \big< (ψ, \partial_u (\phi\cdot ψ)), D^*[f]\big> \epsilon&=
\big<(ψ, \partial_u (\phi\cdot ψ)),(a,b)\big> \epsilon\\
&=\Big(ψ\cdot a+ \partial_u (\phi\cdot ψ)) \cdot b\Big) \epsilon
    \end{split}
\end{equation}
where $a,b$ scalar functions of $f$.\\
\\
Using equation (\ref{eqn:basiceq}) for the wave operator restricted to a null hypersurface, we will find the corresponding $D^*$ and $W$.\\
\\
We consider the following quantities:

\begin{equation}
    A=\big<-\frac{2}{\phi}\partial_v \partial_u (\phi \cdot ψ),f\big> \epsilon=-\frac{2f}{\phi}\partial_v \partial_u (\phi \cdot ψ)dvdμ_{\cancel{g}},
\end{equation}
\\
\begin{equation}
    B=\big<Ω^2 \cancel{Δ}ψ,f\big> \epsilon=fΩ^2 \cancel{Δ}ψdvdμ_{\cancel{g}},
\end{equation}
\\

\begin{equation}
    C=\big<(\cancel{\nabla}Ω^2-2Ω^2 ζ^{\#})\cancel{\nabla}ψ,f\big> \epsilon=f(\cancel{\nabla}Ω^2-2Ω^2 ζ^{\#})\cancel{\nabla}ψdvdμ_{\cancel{g}},
\end{equation}
\\
and
\begin{equation}
    D=\big<wψ,f\big> \epsilon=fwψdvdμ_{\cancel{g}}.
\end{equation}
\\
Then
\begin{equation*}
    \big< D[(ψ, \partial_u (\phi\cdot ψ))],f\big> \epsilon=A+B+C+D.
\end{equation*}
\\
For the quantity A we have 
\begin{equation*}
\begin{split}
    A &=-\frac{2}{\phi}\partial_v \partial_u (\phi \cdot ψ)f dvdμ_{\cancel{g}}\\
    &=- \frac{2}{\phi}\partial_v\partial_u (\phi \cdot ψ)f \cdot \phi^2 dvdμ_{\mathbin{S}^2}\\
    &= - 2\partial_v \partial_u (\phi \cdot ψ)f \cdot \phi dvdμ_{\mathbin{S}^2}\\
    &=\Big(\partial_v(-2\phi f\partial_u (\phi \cdot ψ))+ 2\partial_v (\phi \cdot f) \partial_u (\phi \cdot ψ)\Big) dvdμ_{\mathbin{S}^2}\\
    &=\partial_v(-2\phi f\partial_u (\phi \cdot ψ))dvdμ_{\mathbin{S}^2}+ \frac{2}{\phi^2}\partial_v (\phi \cdot f) \partial_u (\phi \cdot ψ) dvdμ_{\cancel{g}}.
\end{split}
\end{equation*}
Then, for the quantity B we have
\begin{equation*}
    B= Ω^2 \cancel{Δ}ψf dvdμ_{\cancel{g}}= Ω^2 di\mkern-7.5mu/v(\cancel{\nabla}ψ)f dvdμ_{\cancel{g}}.
\end{equation*}
Note that 
\begin{equation*}
\begin{split}
     di\mkern-7.5mu/v(fΩ^2\cancel{\nabla}ψ)&=fΩ^2di\mkern-7.5mu/v(\cancel{\nabla}ψ)+d(fΩ^2)\cdot\cancel{\nabla}ψ= fΩ^2di\mkern-7.5mu/v(\cancel{\nabla}ψ)+f\cancel{\nabla}Ω^2\cdot\cancel{\nabla}ψ +Ω^2\cancel{\nabla}f\cdot\cancel{\nabla}ψ\\
     &\Rightarrow fΩ^2di\mkern-7.5mu/v(\cancel{\nabla}ψ)=di\mkern-7.5mu/v(fΩ^2\cancel{\nabla}ψ)-(f\cancel{\nabla}Ω^2\cdot\cancel{\nabla}ψ +Ω^2\cancel{\nabla}f\cdot\cancel{\nabla}ψ).
\end{split}
\end{equation*}
And therefore,
\begin{equation*}
    B= (\cancel{div}(fΩ^2\cancel{\nabla}ψ)-f\cancel{\nabla}Ω^2\cdot\cancel{\nabla}ψ -Ω^2\cancel{\nabla}f\cdot\cancel{\nabla}ψ) dvdμ_{g\mkern-7.5mu/}.
\end{equation*}
Moreover,
\begin{equation*}
    \begin{split}
       \cancel{div}(fψ\cancel{\nabla}Ω^2)&=d(fψ)\cdot \cancel{\nabla}Ω^2+fψ\cancel{div}(\cancel{\nabla}Ω^2)=f\cancel{\nabla}ψ \cdot \cancel{\nabla}Ω^2+ψ\cancel{\nabla}f \cdot \cancel{\nabla}Ω^2+fψ\cancel{div}(\cancel{\nabla}Ω^2)  \\
       &\Rightarrow f\cancel{\nabla}ψ \cdot \cancel{\nabla}Ω^2= \cancel{div}(fψ\cancel{\nabla}Ω^2)-ψ\cancel{\nabla}f \cdot \cancel{\nabla}Ω^2-fψ\cancel{div}(\cancel{\nabla}Ω^2)
    \end{split}
\end{equation*}
and 
\begin{equation*}
    \begin{split}
     \cancel{div}(Ω^2ψ\cancel{\nabla}f)&=d(Ω^2ψ)\cdot \cancel{\nabla}f+Ω^2ψ\cancel{div}(\cancel{\nabla}f)=Ω^2\cancel{\nabla}ψ \cdot \cancel{\nabla}f+ψ\cancel{\nabla}Ω^2 \cdot \cancel{\nabla}f+Ω^2ψ\cancel{div}(\cancel{\nabla}f)  \\
     &\Rightarrow Ω^2\cancel{\nabla}ψ \cdot \cancel{\nabla}f= \cancel{div}(Ω^2ψ\cancel{\nabla}f)-ψ\cancel{\nabla}Ω^2 \cdot \cancel{\nabla}f-Ω^2ψ\cancel{div}(\cancel{\nabla}f).
    \end{split}
\end{equation*}
So finally, we get that
\begin{equation*}
    B= \cancel{div}(fΩ^2\cancel{\nabla}ψ-fψ\cancel{\nabla}Ω^2-Ω^2ψ\cancel{\nabla}f)dvdμ_{\cancel{g}}+ \bigg( 2ψ\cancel{\nabla}f \cdot \cancel{\nabla}Ω^2+fψ\cancel{div}(\cancel{\nabla}Ω^2)+Ω^2ψ\cancel{div}(\cancel{\nabla}f)\bigg)dvdμ_{\cancel{g}}.
\end{equation*}
\\
For the quantity C we have
\begin{equation*}
    C=(\cancel{\nabla}Ω^2-2Ω^2 ζ^{\#})\cancel{\nabla}ψf dvdμ_{\cancel{g}}.
\end{equation*}
Note  that 
\begin{equation*}
\begin{split}
    \cancel{div}(fψ\cancel{\nabla}Ω^2-fψ2Ω^2 ζ^{\#})&=d(fψ) \cdot \cancel{\nabla}Ω^2 +fψ\cancel{div}(\cancel{\nabla}Ω^2)-d(fψ)\cdot 2Ω^2ζ^{\#}-fψ\cancel{div}(2Ω^2ζ^{\#})\\
    &=fdψ \cdot \cancel{\nabla}Ω^2+ψdf \cdot \cancel{\nabla}Ω^2+fψ\cancel{div}(\cancel{\nabla}Ω^2)\\
    &\;\;\;\;\;-fdψ\cdot 2Ω^2ζ^{\#}-ψdf\cdot 2Ω^2ζ^{\#}-fψ\cancel{div}(2Ω^2ζ^{\#})\\
    &=f\cancel{\nabla}ψ \cdot \cancel{\nabla}Ω^2+ψ\cancel{\nabla}f \cdot \cancel{\nabla}Ω^2+fψ\cancel{div}(\cancel{\nabla}Ω^2)\\
    &\;\;\;\;\;-f\cancel{\nabla}ψ\cdot 2Ω^2ζ^{\#}-ψ\cancel{\nabla}f\cdot 2Ω^2ζ^{\#}-fψ\cancel{div}(2Ω^2ζ^{\#}).
\end{split}
\end{equation*}
And thus,
\begin{equation*}
    \begin{split}
        C&= \cancel{div}(fψ\cancel{\nabla}Ω^2-fψ2Ω^2 ζ^{\#})dvdμ_{\cancel{g}}- fψ\cancel{div}(\cancel{\nabla}Ω^2) dvdμ_{\cancel{g}}\\
        &- ψ\cancel{\nabla}f \cdot \cancel{\nabla}Ω^2 dvdμ_{\cancel{g}}+ ψ\cancel{\nabla}f\cdot 2Ω^2ζ^{\#} dvdμ_{\cancel{g}}+ fψ\cancel{div}(2Ω^2ζ^{\#}) dvdμ_{\cancel{g}}.
    \end{split}
\end{equation*}
\\
So finally, C is equal to 
\begin{equation*}
  C=\Big(\cancel{div}(fψ\cancel{\nabla}Ω^2-fψ2Ω^2 ζ^{\#})- fψ(\cancel{div}(\cancel{\nabla}Ω^2 -2Ω^2ζ^{\#} )- ψ\cancel{\nabla}f \cdot (\cancel{\nabla}Ω^2-2Ω^2ζ^{\#})\Big) dvdμ_{\cancel{g}}.
\end{equation*}
\\
Thus, we find
$$\big<D[(ψ, \partial_u (\phi\cdot ψ))], f\big> dvdμ_{\cancel{g}}=A+B+C+D$$
$$=\partial_v(-2\phi f\partial_u (\phi \cdot ψ))dvdμ_{S^2}+\Big( \frac{2}{\phi^2}\partial_v (\phi \cdot f) \partial_u (\phi \cdot ψ) +\cancel{div}(fΩ^2\cancel{\nabla}ψ-fψ\cancel{\nabla}Ω^2-Ω^2ψ\cancel{\nabla}f)$$
$$+ 2ψ\cancel{\nabla}f \cdot \cancel{\nabla}Ω^2+fψ\cancel{div}(\cancel{\nabla}Ω^2)+Ω^2ψ\cancel{div}(\cancel{\nabla}f)+\cancel{div}(fψ\cancel{\nabla}Ω^2-fψ2Ω^2 ζ^{\#})$$
$$-fψ(\cancel{div}(\cancel{\nabla}Ω^2 -2Ω^2ζ^{\#} )- ψ\cancel{\nabla}f \cdot (\cancel{\nabla}Ω^2-2Ω^2ζ^{\#})+wψ f \Big)dvdμ_{\cancel{g}}$$
\\
which leads to
\begin{equation*}
    \begin{split}
        \big<D[(ψ, \partial_u (\phi\cdot ψ))], f\big> dvdμ_{\cancel{g}}&=\partial_v(-2\phi f\partial_u (\phi \cdot ψ))dvdμ_{S^2}\\
&\;\;\;\;+(\cancel{div}(fΩ^2\cancel{\nabla}ψ-Ω^2ψ\cancel{\nabla}f-fψ2Ω^2 ζ^{\#}))dvdμ_{\cancel{g}}\\
&\;\;\;\;+\Big( \frac{2}{ϕ^2}\partial_v (\phi \cdot f) \partial_u (\phi \cdot ψ) +Ω^2ψ\cancel{div}(\cancel{\nabla}f)\\
&\;\;\;\;+ψ\cancel{\nabla}f(\cancel{\nabla}Ω^2 +2Ω^2ζ^{\#})+fψ\big(\cancel{div}(2Ω^2 ζ^{\#})+w\big)\Big)dvdμ_{\cancel{g}}.
    \end{split}
\end{equation*}
\vspace{0.1cm}
\\
Using the relation above, we find
\begin{equation}
\begin{split}
    \big< (ψ,\partial_u (\phi \cdot ψ)), D^*[f]\big> \epsilon &= \Big( \frac{2}{{\phi}^2}\partial_v (\phi \cdot f) \partial_u (\phi \cdot ψ)\Big)dvdμ_{\cancel{g}}\\
    & +\Big(Ω^2ψ\cancel{div}(\cancel{\nabla}f)+ψ\cancel{\nabla}f(\cancel{\nabla}Ω^2 +2Ω^2ζ^{\#})+fψ\big(\cancel{div}(2Ω^2 ζ^{\#})+w\big)\Big)dvdμ_{\cancel{g}}.
\end{split}
\end{equation}
\\
Therefore, for the associate adjoint $D^*$, which takes a function $f$ into a pair of functions
\begin{equation*}
    D^* [f]=(a,b),
\end{equation*}
the functions $a$ and $b$ are equal to
\begin{equation*}
    a=Ω^2\cancel{Δ}f+ (\cancel{\nabla}Ω^2+2Ω^2ζ^{\#})\cdot\cancel{\nabla}f +(\cancel{div}(2Ω^2ζ^{\#})+w)f
\end{equation*}
and
\begin{equation*}
    b=\frac{2}{ϕ^2}\partial_v (ϕ \cdot f).
\end{equation*}
Thus,
\begin{equation}
   D^* [f]=\bigg(Ω^2\cancel{Δ}f+ (\cancel{\nabla}Ω^2+2Ω^2ζ^{\#})\cancel{\nabla}f +(\cancel{div}(2Ω^2ζ^{\#})+w)f,\frac{2}{\phi^2}\partial_v (\phi \cdot f)\bigg).
\end{equation}
\\
\\
Finally, we find that
\begin{equation}
\label{eqn:dW}
    \boxed{dW[ψ,\partial_u (\phi\cdot ψ),f]=\partial_v(-2\phi f\partial_u (\phi \cdot ψ))dvdμ_{\mathbin{S}^2}
+(\cancel{div}(fΩ^2\cancel{\nabla}ψ-Ω^2ψ\cancel{\nabla}f-fψ2Ω^2 ζ^{\#}))dvdμ_{\cancel{g}}}
\end{equation}
\\
To find $W$, we will have to bring the right hand side of equation (\ref{eqn:dW}) in the form of an exterior derivative. For simplicity in calculations we define
\begin{equation}
    h=-2\phi f\partial_u (\phi \cdot ψ)
\end{equation}
and
\begin{equation}
    X=fΩ^2\cancel{\nabla}ψ-Ω^2ψ\cancel{\nabla}f-fψ2Ω^2 ζ^{\#}.
\end{equation}
\\
Then $dW$ can be written as

\begin{equation}
    dW[ψ,\partial_u (\phi\cdot ψ),f]=\partial_v(h)dvdμ_{\mathbin{S}^2}
+(\cancel{div}(X))dvdμ_{\cancel{g}}.
\end{equation}
\\
By the definition of the exterior derivative $d$, we write
\begin{equation*}
    d(hdμ_{\mathbin{S}^2})=\partial_v(h)dv \wedge dμ_{\mathbin{S}^2}
\end{equation*}
and thus, 
\begin{equation*}
  d(hdμ_{S^2})=d\big(\frac{h}{\phi^2}dμ_{\cancel{g}}\big).  
\end{equation*}
\\
Also, using the volume form of the spheres $\cancel{\epsilon}=\sqrt{\cancel{g}} d\theta^1 \wedge d\theta^2$, we define the volume form of the null hypersurface to be  $\epsilon$ given by $\epsilon=dv \wedge \cancel{\epsilon}$ .\\
\\
Below we present two lemmas that will be needed in order to express (\ref{eqn:dW}) as an exterior derivative.\\
\\
\textbf{Lemma 1}: Let $X$ be a vector field tangential to the sections $S_v$. Then, $\cancel{d}(\cancel{\epsilon} \iprod X) \wedge dv=d(\cancel{\epsilon} \iprod X) \wedge dv$.
\begin{proof} 

Let $X=X^1\partial_{\theta^1} + X^2\partial_{\theta^2}$ and $\cancel{\epsilon}=\sqrt{\cancel{g}}d\theta^1 \wedge d\theta^2$ the volume form of the sphere. Then, 
$$\cancel{\epsilon} \iprod X=i_X \cancel{\epsilon}=\sqrt{\cancel{g}} (X^1d\theta^2-X^2d\theta^1) $$
Let's call
$$\sqrt{\cancel{g}} (X^1d\theta^2-X^2d\theta^1)=C_id\theta^i $$
where summation over $i$ is implied and $C_1=-\sqrt{\cancel{g}}X^2$, $C_2=\sqrt{\cancel{g}}X^1$.\\
\\
Then, we can write
$$\cancel{\epsilon} \iprod X=C_id\theta^i $$
and by applying the angular exterior derivative we find
$$\cancel{d}(\cancel{\epsilon} \iprod X)=\cancel{d}(C_id\theta^i)=\partial_jC_id\theta^j \wedge d\theta^i$$
where again summation over $i$ and $j$ is implied.
Similarly, taking the exterior derivative on $\mathcal{H}$ we find
\begin{equation*}
    \begin{split}
     &d(\cancel{\epsilon} \iprod X)=d(C_id\theta^i)=\partial_vC_idv \wedge d\theta^i + \partial_jC_id\theta^j \wedge d\theta^i   \\
     \\
     & \Rightarrow d(\cancel{\epsilon} \iprod X)=\partial_vC_idv \wedge d\theta^i + \cancel{d}(\cancel{\epsilon} \iprod X)
    \end{split}
\end{equation*}
and thus
\begin{align}
    \begin{split}
d(\cancel{\epsilon} \iprod X) \wedge dv
 & = \partial_vC_idv \wedge d\theta^i \wedge dv + \cancel{d}(\cancel{\epsilon} \iprod X) \wedge dv\\
 & =\cancel{d}(\cancel{\epsilon} \iprod X) \wedge dv
\end{split}
\end{align}
which proves our claim.\\
\qedhere
\end{proof}

\newpage

\textbf{Lemma 2:} Let $X$ be a vector field tangential to the sections $S_v$. Then, $\cancel{div}(X))dvdμ_{\cancel{g}}=\cancel{div}(X))\epsilon=\cancel{div}(X))dv \wedge \cancel{\epsilon}=d(\cancel{\epsilon} \iprod X)\wedge dv=d(\epsilon \iprod X)$.\\
\begin{proof}

Let $\epsilon=dv \wedge \cancel{\epsilon}= dv \wedge dμ_{\cancel{g}}=\sqrt{\cancel{g}} dv \wedge d\theta_1 \wedge d\theta_2$ be the volume form on $\mathcal{H}$. Then, we have
$$\cancel{div}(X))dvdμ_{\cancel{g}}=\cancel{div}(X))\epsilon=\cancel{div}(X))dv \wedge \cancel{\epsilon}=dv \wedge \cancel{div}(X))
\cancel{\epsilon}.$$
Using the formula that connects the divergence with the exterior derivative we can write
$$\cancel{div}(X))
\cancel{\epsilon}=\cancel{d}(\cancel{\epsilon} \iprod X).$$
\\
Lemma 1 shows that $\cancel{d}(\cancel{\epsilon} \iprod X) \wedge dv=d(\cancel{\epsilon} \iprod X) \wedge dv$. Thus, we are left to show that
\begin{equation*}
    d(\cancel{\epsilon} \iprod X) \wedge dv=d(\epsilon \iprod X).
\end{equation*}
We showed that for a vector field $X$ such that $X=X^1\partial_{\theta^1} + X^2\partial_{\theta^2}$, we have
\begin{equation*}
    \cancel{\epsilon}  \iprod X=C_id\theta^i 
\end{equation*}
where $\sqrt{\cancel{g}} (X^1d\theta^2-X^2d\theta^1)=C_id\theta^i $. 
This yields
\begin{equation} \label{eq1}
\begin{split}
d(\cancel{\epsilon} \iprod X) \wedge dv & = \partial_jC_id\theta^j \wedge d\theta^i\wedge dv \\
 & = \partial_{\theta^1}(\sqrt{\cancel{g}} X^1)d\theta^1 \wedge d\theta^2 \wedge dv-\partial_{\theta^2}(\sqrt{\cancel{g}} X^2)d\theta^2 \wedge d\theta^1 \wedge dv+ \partial_v (0) dv \wedge dv\\
 & = \partial_{\theta^1}(\sqrt{\cancel{g}} X^1)d\theta^1 \wedge d\theta^2 \wedge dv-\partial_{\theta^2}(\sqrt{\cancel{g}} X^2)d\theta^2 \wedge d\theta^1 \wedge dv\\
 & = \partial_{\theta^1}(\sqrt{\cancel{g}} X^1)d\theta^1 \wedge d\theta^2 \wedge dv+\partial_{\theta^2}(\sqrt{\cancel{g}} X^2)d\theta^1 \wedge d\theta^2 \wedge dv\\
 & = (\partial_{\theta^1}(\sqrt{\cancel{g}} X^1)+\partial_{\theta^2}(\sqrt{\cancel{g}} X^2))d\theta^1 \wedge d\theta^2 \wedge dv\\
 & = (\partial_{\theta^1}(\sqrt{\cancel{g}} X^1)+\partial_{\theta^2}(\sqrt{\cancel{g}} X^2)) dv\wedge d\theta^1 \wedge d\theta^2 \\
 & = (\partial_{\theta^1}(\sqrt{\cancel{g}} X^1)+\partial_{\theta^2}(\sqrt{\cancel{g}} X^2))\epsilon.
\end{split}
\end{equation}

The next step is to compute $d(\epsilon \iprod X)$. We start with $\epsilon \iprod X$ as shown below
\begin{equation*}
    \epsilon \iprod X = i_X \epsilon=X^1\sqrt{\cancel{g}} d\theta^2 \wedge dv-X^2 \sqrt{\cancel{g}} d\theta^1 \wedge dv +0d\theta^1 \wedge d\theta^2=X^1\sqrt{\cancel{g}} d\theta^2 \wedge dv-X^2 \sqrt{\cancel{g}} d\theta^1 \wedge dv.
\end{equation*}
If we take the exterior derivative of this expression we get
\begin{equation} \label{eq2}
\begin{split}
d(\epsilon \iprod X)
 & = \partial_{\theta^1}(\sqrt{\cancel{g}}X^1)d{\theta^1} \wedge d{\theta^2} \wedge dv-\partial_{\theta^2}(\sqrt{\cancel{g}}X^2 ) d{\theta^2} \wedge d{\theta^1} \wedge dv\\
 & = (\partial_{\theta^1}(\sqrt{\cancel{g}}X^1)+\partial_{\theta^2}(\sqrt{\cancel{g}}X^2 ) )d{\theta^1} \wedge d{\theta^2} \wedge dv\\
 & = (\partial_{\theta^1}(\sqrt{\cancel{g}}X^1)+\partial_{\theta^2}(\sqrt{\cancel{g}}X^2 ) )dv \wedge d{\theta^1} \wedge d\theta^2\\
 & = (\partial_{\theta^1}(\sqrt{\cancel{g}}X^1)+\partial_{\theta^2}(\sqrt{\cancel{g}}X^2) )\epsilon.
\end{split}
\end{equation}
\\
Therefore, we showed that 
\begin{equation*}
    d(\cancel{\epsilon} \iprod X) \wedge dv=d(\epsilon \iprod X).
\end{equation*}
 \\
 And thus, we proved that
 \begin{equation*}
     \cancel{div}(X))dvdμ_{\cancel{g}}=d(\epsilon \iprod X).
 \end{equation*}
\end{proof}

Finally, we can write 
\begin{equation*}
    dW[ψ,\partial_u (\phi\cdot ψ),f]=d\big(\frac{h}{\phi^2}\cancel{\epsilon}\big)+d(\epsilon \iprod X)
\end{equation*}
\begin{equation*}
    dW[ψ,\partial_u (\phi\cdot ψ),f]=d\big(\frac{h}{\phi^2}\cancel{\epsilon}+\epsilon \iprod X\big)
\end{equation*}
which gives us $W$, 
\begin{equation}
    W=\frac{h}{\phi^2}\cancel{\epsilon}+\epsilon \iprod X.
\end{equation}
\\
By the definition of the associate adjoint operator as given in (\ref{eqn:basiceq}), if $ψ$ is a solution of the wave operator restricted on $\mathcal{H}$ such that $D[(ψ, \partial_u (\phi\cdot ψ))]=0$ and $f$ is a function that satisfies $D^*[f]=0$, then $dW[ψ, \partial_u (\phi\cdot ψ)]=0$.\\
\\
Evoking the Stoke's theorem on forms
\begin{equation}
    \int _{H} dW = \int _{\partial H} W
\end{equation}
\\
we find that in the case where $dW=0$
\begin{equation}
  \int _{\partial H} W=0.
\end{equation}
In the case we examine, the boundary of $H$ is spheres, so the restriction of $W$ on $\partial H$ will be
\begin{equation*}
    W|_{\partial H}= \frac{h}{\phi^2}\cancel{\epsilon}|_{\partial H}+\epsilon \iprod X|_{\partial H}=\frac{h}{\phi^2}\cancel{\epsilon}|_{\partial H}=\frac{h}{\phi^2}\cancel{\epsilon}
\end{equation*}
since $\epsilon \iprod X=B_id\theta^i \wedge dv$ for some coefficients $B_i$ where $i=\{1,2\}$.\\
Substitution of $h$ back to this equation yields

\begin{equation}
      \int _{\partial H}\frac{-2\phi f\partial_u (\phi \cdot ψ)}{\phi^2}\cancel{\epsilon}=\int _{S_v}-2\phi f\partial_u (\phi \cdot ψ)dμ_{\mathbin{S}^2}-\int _{S_0}-2\phi f\partial_u (\phi \cdot ψ)dμ_{\mathbin{S}^2}=0
\end{equation}
giving us that the quantity

\begin{equation}
  \int _{S_v}-2\phi f\partial_u (\phi \cdot ψ)dμ_{\mathbin{S}^2}
\end{equation}
is conserved because this holds for every $v$.





\newpage
\nocite{*}
\bibliographystyle{IEEEannot}
\bibliography{annot}

\end{document}